\documentclass[aps,prf,prfprint,superscriptaddress,amsmath,amssymb,onecolumn,longbibliography]{revtex4-2}
\usepackage[utf8]{inputenc}
\usepackage[T1]{fontenc}
\usepackage{graphicx}
\usepackage[usenames]{color}
\usepackage{amsmath,amssymb}
\usepackage{gensymb}
\usepackage{hyperref}
 \usepackage{cancel}

\usepackage[normalem]{ulem}

\newcommand\add[1]{\textcolor{black}{#1}}
\newcommand\DEL[1]{}  
\newcommand{\rep}[2]{\textcolor{black}{\sout{}}\textcolor{black}{#2}}

\begin{document}
\title{Fluid-particle suspension by gas release from a granular bed}

\author{Tess Homan}
\affiliation{Univ Lyon, ENS de Lyon, Univ Claude Bernard, CNRS, Laboratoire de Physique, Lyon, France}
\affiliation{Mechanical Engineering, Eindhoven University of Technology, Eindhoven, The Netherlands}
\author{Val\'erie Vidal}
\affiliation{Univ Lyon, ENS de Lyon, Univ Claude Bernard, CNRS, Laboratoire de Physique, Lyon, France}
\author{Cl\'ement Picard}
\affiliation{Univ Lyon, ENS de Lyon, Univ Claude Bernard, CNRS, Laboratoire de Physique, Lyon, France}
\author{Sylvain Joubaud}
\affiliation{Univ Lyon, ENS de Lyon, Univ Claude Bernard, CNRS, Laboratoire de Physique, Lyon, France}
\affiliation{Institut Universitaire de France (IUF), 1 rue Descartes, 75005 Paris, France}
\date{\today} 

\begin{abstract}
{We have studied experimentally particle suspension when injecting a gas at the bottom of an immersed granular layer confined in a Hele-Shaw cell. This work focuses on the dynamics of particles slightly denser than the surrounding fluid. The gas, injected at constant flow-rate, rises through the granular bed and then forms bubbles that entrain particles in the above liquid layer. The particles settle down on the edges of the cell, avalanche on the crater formed at the granular bed free surface, and are further entrained by the continuous bubbling at the center. We report the existence of a stationary state, resulting from the competition between particle entrainment and sedimentation. The average \rep{packing fraction}{solid fraction} in the suspension is derived from a simple measurement of the granular bed apparent area. A phenomenological model based on the balance between particle lift by bubbles at the center of the cell and their settling on its sides demonstrates that most of the particles entrained by bubbles come from a global recirculation of the suspension.}
\end{abstract}

\maketitle

\section{Introduction}
{Recently, there has been a growing research interest for multiphase flows, as their understanding is one of the grand natural and industrial challenges in fluid dynamics~\cite{DauxoisLesHouches2019}. Among the multitude of geophysical flows, gas release in a particle-laden fluid is a widespread phenomenon that may have drastic consequences on the environment~\cite{Olsen2016}. On the one hand, the understanding of methane production and transport in sedimentary basins and its subsequent release is crucial in terms of climate change and global warming~\cite{Judd2002,Svensen2004,Naudts2008}. On the other hand, exsolved volatiles rising through crystal-rich magmas strongly influence volcanic eruption dynamics~\cite{Sable2006,Oppenheimer2015,Cashman2017}. Quantifying the mechanisms leading to such resuspension and the generated turbidity current is also essential for the effects of human activities from the production of crude oil from the Canadian oil sands~\cite{Lawrence2016} and deep-sea mining~\cite{Peacock2018}. In industry, catalytic gas-fluidized bed reactors have been widely investigated for the optimization of chemical processes~\cite{Dudukovic2002,Rados2003,Pangarkar2008}. In most of these applications, the interplay between the gas and the particles is one of the key parameters of the global dynamics of such multiphase flows. Therefore, understanding and quantifying the ability of gas to entrain and maintain particles in suspension is a question of paramount importance.}

{To tackle this question, we consider particle entrainment immersed granular beds. The resuspension of particles forming this solid-like settled state and the induced erosion, chimney or crater formation have been investigated using different mechanisms such as shearing flow~\cite{Aussilous2013,Houssais2016}, the impact of liquid jets~\cite{Badr2014,Sutherland2014,Badr2016,Vessaire2020}, thermal convection or plume emission~\cite{Martin1988,Solomatov1993,Lavorel2009,Morize2017,Herbert2018}, underground cavity collapse~\cite{Loranca2015}, fluidization~\cite{Houssais2019}, etc. The present paper focuses on gas release from a granular bed, a scenario highlighted in the above applications. In the past years, two classes of model systems have been developed  to exhibit the physical mechanisms at stake in such three-phase flows, where the coupling between the grains, gas and liquid may have a strong impact on the global dynamics. On the one hand, to remove or neglect the effect of gravity, experiments have been performed in horizontal setups and/or using isodense particles~\cite{Johnsen2006,Chevalier2007,Johnsen2008,Cheng2008,Chevalier2009,Sandnes2011,Holtzman2012, Hooshyar2013, Madec2020}. On the other hand, buoyancy-driven systems have mostly focused on gas patterns in a dense granular bed, with particles much heavier than the surrounding fluid~\cite{Gostiaux2002,Geistlinger2006,Kong2009,Varas2009,Kong2010,VarasJAN2011,Varas2013,Varas2015}. Depending on the gas injection flow rate or pressure and the local \rep{packing fraction}{solid fraction}, the gas may either percolate through the grains or fracture the bed. At the grain free surface, the successive ejection of gas bubbles entrains particles in the liquid. The competition between the particle lift and sedimentation leads at long time to a crater formation~\cite{Varas2009}.}

In the present paper, we study experimentally the global characteristics of the suspension formed by particles slightly heavier than the surrounding fluid, which are entrained by continuous gas injection. In particular, we focus on the balance between entrainment by the bubble rise and sedimentation.  The goal is to identify and quantify the controlling parameters of the extension and average \rep{packing fraction}{solid fraction} of the suspension. We demonstrate that this latter can be estimated at each time from the size of the granular bed which remains at the cell bottom. We then quantify the existence and properties of a steady state when varying the cell geometry and particles properties. We finally propose a simple model based on the balance between entrainment and sedimentation, which unravels the particule entrainment mechanism in the stationary state.

The paper is organized as follows. After a description of the experimental set-up (section~\ref{sec:expsetup}), the analysis of the experimental results and the influence of the various parameters is presented in section~\ref{sec:exp_results}. In a second stage, the different ingredients of a simple model are presented in \add{section~\ref{sec:model}} and its \rep{prediction}{predictions} compared to the experimental results in section~\ref{sec:Particle_susp}. Finally, we conclude and draw some perspectives in section~\ref{sec:conclusion}. 

\section{Experimental setup}
\label{sec:expsetup}

\begin{figure}[t!]
\includegraphics[width=0.8\columnwidth]{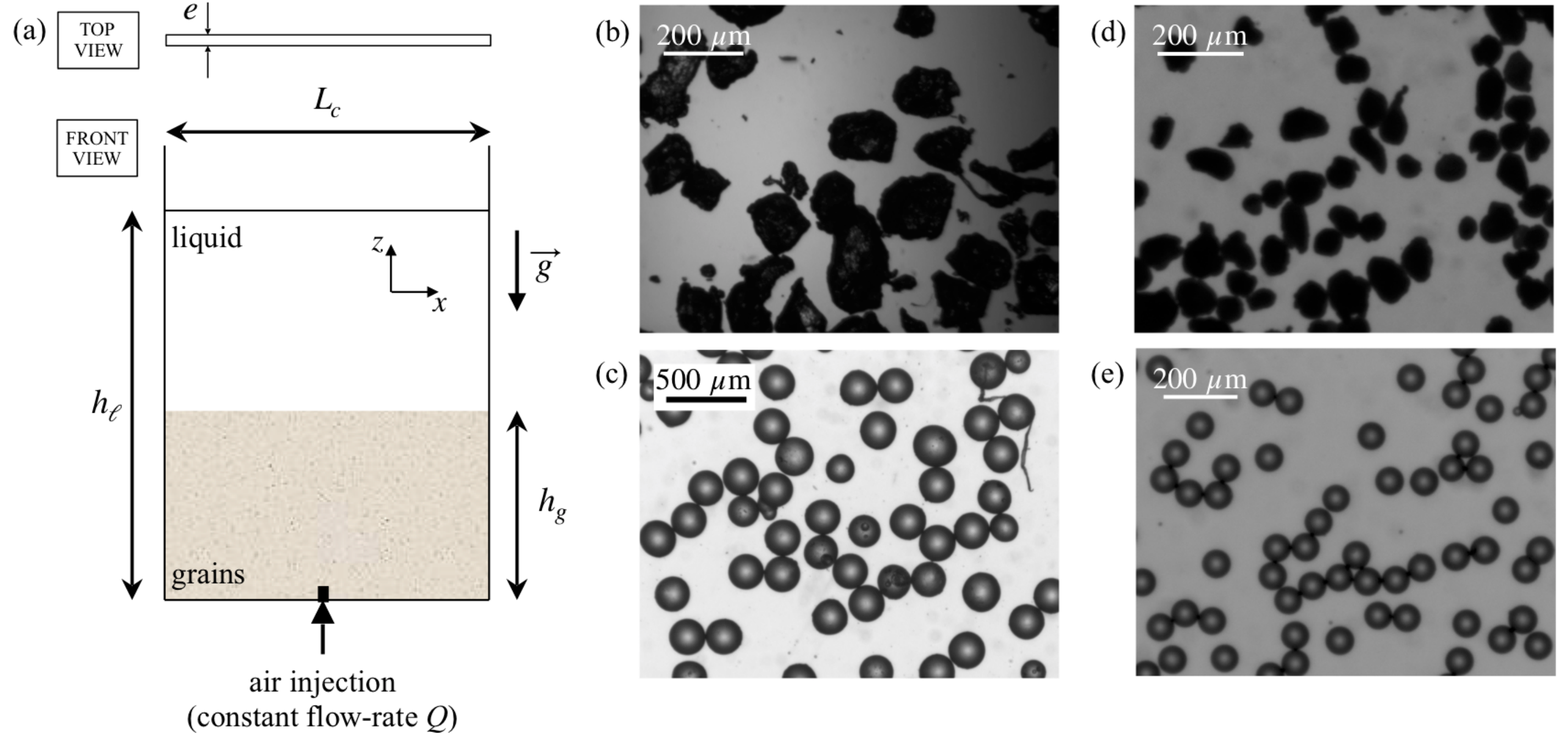}
\caption{
(a) Schematic view of the experimental setup. Air is injected at constant flow rate $Q$ at the bottom of an immersed granular layer in a Hele-Shaw cell (see text). The subsequent suspension and the remaining granular bed are observed using shadowgraphy. The different notations of the geometrical parameters are indicated.
On the right, images of different batches of grains~(see~Table~\ref{tab:grains}): (b) non spherical polydisperse polystyrene beads (PS 130P); (c) spherical monodisperse polystyrene beads (PS 250M); (d)  non spherical polydisperse PVC beads (PVC 110P); (e) spherical monodisperse PS beads (PS 80M).}
\label{fig:expsetup}
\end{figure}

The experimental setup, sketched in figure~\ref{fig:expsetup}(a), consists of a vertical Hele-Shaw cell of height 30~cm, width $L_{\textrm{c}}$ ($L_{\textrm{c}}=13.6, 24$ or $35.6$~cm) and gap $e$ ($e= 2$ or $3$~mm). The cell is filled with ethanol (absolute, Merck Millipore, density $\rho_\ell = 789$~kg/m$^{3}$, viscosity $\mu=1.2 \times 10^{-3}$~Pa.s) and beads of either polystyrene (PS, density $\rho_g=1059$~kg/m$^{3}$) or polyvinyl chloride (PVC, density $\rho_g=1379$~kg/m$^{3}$), with different sizes and shapes (typical diameter $d$, monodisperse (M) or polydisperse (P), see Table~\ref{tab:grains}). Images of the different batches are displayed on figure~\ref{fig:expsetup}\rep{c-f}{b-e}. Note that the use of ethanol prevents the formation of particles aggregates. Air is injected at constant flow-rate $Q$ at the bottom of the cell through a central gas-inlet (inner diameter $1$~mm). The flow-rate is varied between $Q=0.013$ and $1.5$~L/min by means of a mass-flow controller (Bronkhorst, Mass Stream D-5111 for $0.01\leq Q \leq 0.05$~L/min and D-6311 for $0.05\leq Q \leq 2$~L/min). A reproducible initial condition is obtained by mixing the particles and the liquid with a strong air flow-rate ($2$~L/min) for $3$~minutes. The air flow is then turned off and the particles left to sediment gently, leading to a homogeneous loose-packing initial bed. As expected, the obtained \rep{packing fraction}{solid fraction}, $\varphi_b^0$ for the monodisperse spherical particles corresponds to a random loosely packed state~\cite{AndreottiB2013} (Table~\ref{tab:grains}). Note that $\varphi_b^0$ is much smaller for polydisperse non-spherical angular particles, for which the bed is in a very loose state~\cite{Brouwers2006} (Table~\ref{tab:grains}). 
The initial bed height, $h_g$, is varied up to 10~cm and the liquid height is ajusted to a given value $h_\ell>h_g$ (see Fig.~\ref{fig:expsetup}(a)). The ratio $h_\ell/h_g$ lies in the range 1.2--4.

\begin{table*}
\begin{center}
\begin{tabular}{c|c|c|c|c|c|c|c|c}
\hline \hline
~particles~ & ~provider~ & ~shape~ & ~distribution~ & ~$\Delta\rho$ [kg/m$^{3}$]~ & ~$d$ [$\mu$m]~ & ~$\varphi_b^0$ [\%]~ & \add{$U_s$} [mm/s]~&~\add{$\textrm{Ar}$}\\
\hline \hline
PVC 110P & Goodfellow$^\circledR$ & non spherical & polydisperse & 590 & $110 \pm 50$& ~$41.9 \pm 0.5$~& 3.2 ~&~\add{0.23}\\
PS 130P    & Goodfellow$^\circledR$  & non spherical & polydisperse  & 270 & $130 \pm 80$  & $42.0 \pm 0.5 $&  2.1 ~&~\add{0.18}\\
PS 250M  & Dynoseeds$^\circledR$   & spherical & monodisperse & 270 & $230 \pm 10$  &  $56.2 \pm 0.5 $& 6.5 ~&~\add{0.98}\\
PS 80M    & Dynoseeds$^\circledR$   & spherical & monodisperse & 270 & $80 \pm 5$ &  $57.8 \pm 0.5$& 0.8 ~&~\add{0.04}\\
\hline \hline 
\end{tabular}
\caption{\label{tab:grains}
Characteristics of the polystyrene (PS) or polyvinyl chloride (PVC) beads used in the experiments. $\Delta \rho = \rho_g - \rho_\ell$ is the density difference between the particles and the fluid, $d$ the typical particle diameter, and $\varphi_b^0$ the initial bed \rep{packing fraction}{solid fraction} (see text). \rep{$U_s= \frac{\Delta \rho g d^2}{18 \mu}$}{$U_s= \Delta \rho g d^2/(18 \mu)$} is the Stokes (settling) velocity of a single particle of typical diameter $d$ in a fluid of viscosity $\mu$ (see section~\ref{sec:model}). \add{$\textrm{Ar}=\rho_\ell U_s d/\mu$ is the  Archimedes number, which
corresponds to the particle Reynolds number based on the Stokes velocity $U_s$.}}
\end{center}
\end{table*}

The setup is illuminated from behind by a strong homogeneous backlight (Dalle LED, Euroshopled). Shadowgraph imaging of the container, the granular bed and the suspension is performed using a camera (PixeLINK, PL-B741U) capturing images at~1~Hz. 
\rep{Finally, a}{A} contour detection, based on intensity thresholding, makes it possible to infer the granular bed area $A$, and thus its volume $Ae$, for each image, and to get its temporal evolution during each experiment. We denote $A_0$ the initial bed area \add{and $h_g=A_0/L_c$ the initial bed height.} \add{Finally, bubble contour detection is performed in the suspension in the stationary regime, to quantify the typical bubble size (see section~\ref{sec:LbUb}).}

\section{Experimental results}
\label{sec:exp_results}

\subsection{Phenomenology}

\begin{figure}[t]
\includegraphics[width=\linewidth]{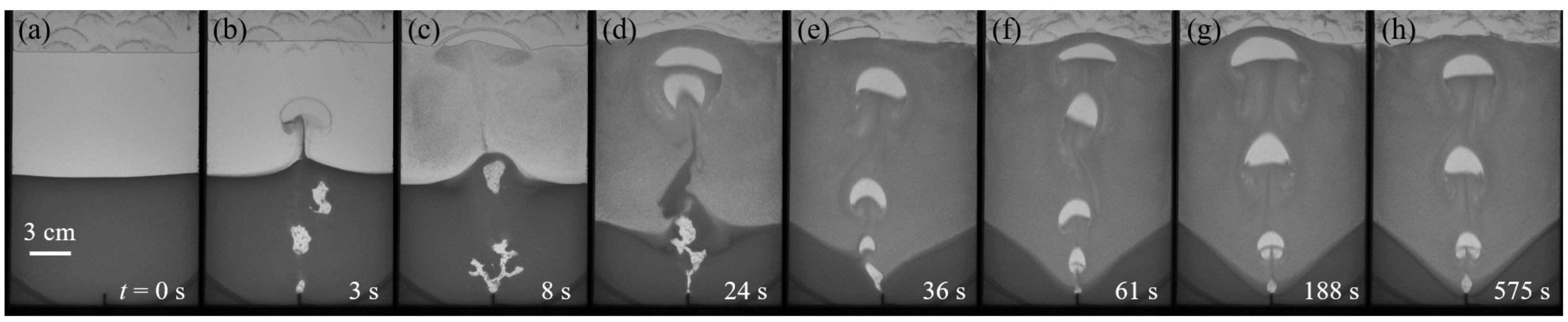}
\caption{  \label{fig:TempEvol}
Temporal evolution of the bed and suspension [PS 130P, $Q=200$~mL/min, $h_g = 10\pm 0.2$~cm, $h_\ell = 20\pm0.2$~cm, $L_{\textrm{c}}=13.6$~cm\add{, $e=2$~mm}]. (a) Initial loosely packed bed. (b)-(c) After turning on the gas injection, air rises through the granular bed and forms bubbles which entrain particles in their wake in the above liquid layer. (d)-(e) A crater grows and the suspension becomes denser. (f-h) The system reaches a stationary state in which the volume of the granular bed and the average \rep{packing fraction}{solid fraction} of the suspension remain constant.}
\end{figure}
%

\begin{figure}[h]
\includegraphics[width=0.8\linewidth]{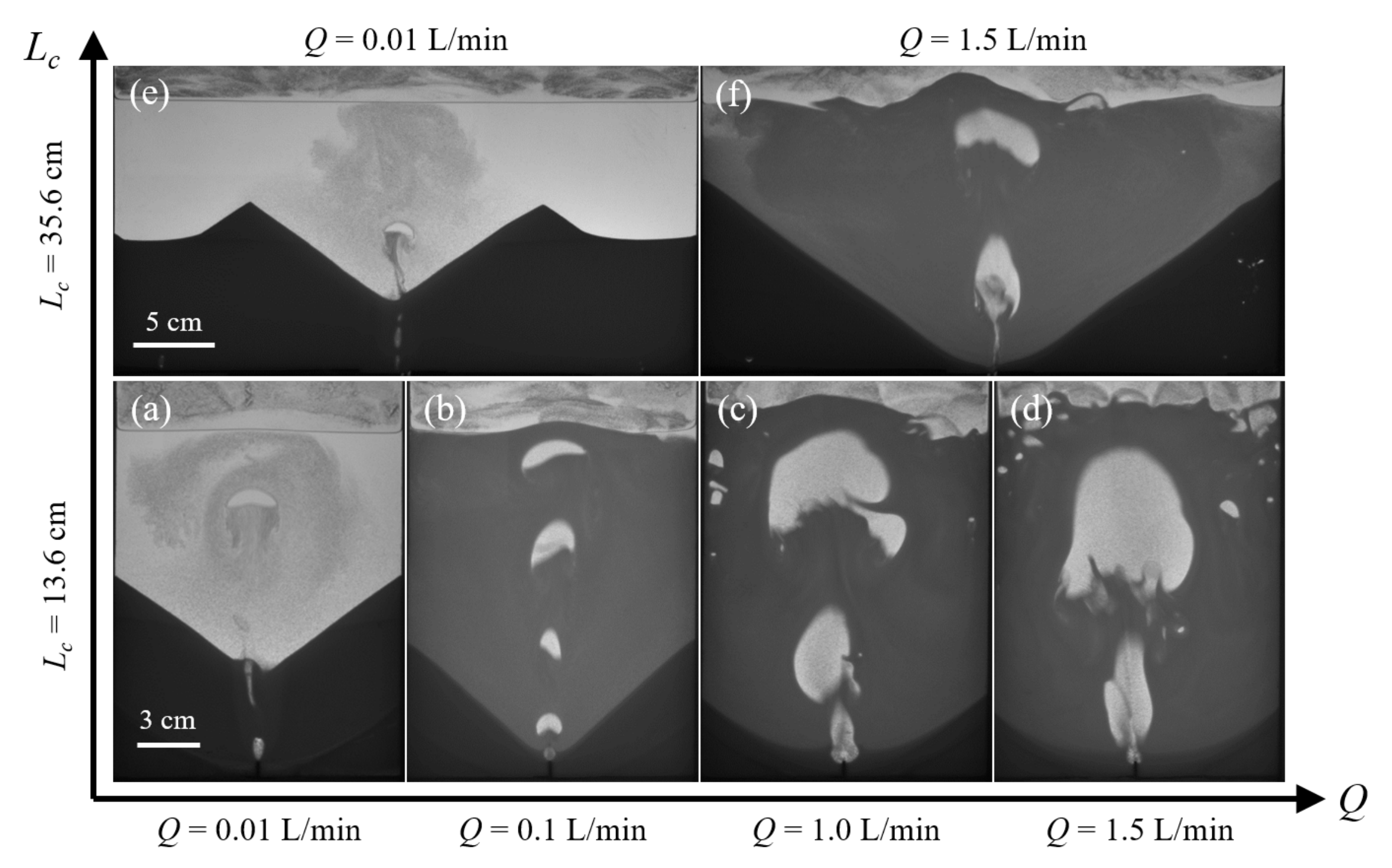}
\caption{ \label{fig:LcQdiagram}
Snapshots of the experiment in the stationary regime for different values of the flow rate $Q$ and different cell width $L_c$ [PVC 110P, $h_g = 9\pm 0.5$~cm, $h_\ell = 18\pm0.5$~cm]. \add{The thickness of the cell is $e=2$~mm for the lower panel [$L_c=13.6$~cm] and $e=3$~mm for the upper panel [$L_c=35.6$~cm].}}
\end{figure}

Figure~\ref{fig:TempEvol} displays a typical experiment, where a time lapse shows the evolution of the granular bed and the suspended particles. At $t~=~0$~s, air is injected at constant flow rate $Q$ at the bottom of the immersed granular bed. The gas initially invades the granular bed, then air bubbles escape and rise through the above liquid layer, entraining particles in their wake [Figs.~\ref{fig:TempEvol}(b)-(c)]. This process leads to the formation of a crater which size increases in time [Figs.~\ref{fig:TempEvol}(c)-(f)]. When the particles deposit on the inner part of the crater, they avalanche back to the center and are further entrained by the continuous gas injection. The system finally reaches a stationary state (Fig.~\ref{fig:TempEvol}(f-h)) characterized by a \textit{suspension}, resulting from the balance between particles lifted by gas bubbles and sedimentation; and a {\it granular bed}, corresponding to the particles that are not entrained by the gas flow. Note that a small transition region exists between the granular bed and the suspension, corresponding to the avalanching particles, slightly less dense than the granular bed. Its size is at most a few percent of the granular bed's size and is included in the bed area, $A$, by our thresholding method.

To carefully investigate the existence and the characteristics of this stationary state, experiments have been performed for a wide range of experimental parameters. Typical snapshots are shown in figure~\ref{fig:LcQdiagram} for different flow rates $Q$ and cell widths $L_c$.  At constant $L_c$, when increasing $Q$, we observe that the area occupied by the granular bed decreases and the suspension becomes darker, confirming that its volume fraction increases. In addition, the typical width occupied by the suspension, denoted as $L_s$, increases until it reaches the lateral boundaries of the cell, $L_c$. In most of our experiments, $L_s \simeq L_c$. \add{In this work, except when explicitely mentionned, we will focus on this configuration only}. While we retrieve the typical crater shape with two dunes reported by~\cite{Varas2009} for small $Q$ and large $L_c$ (top-left snapshot in figure~\ref{fig:LcQdiagram}), the flank of the crater is limited by the cell boundary \add{for} large $Q$ and/or small $L_c$. 

The crater and suspension characteristics in the stationary state do not depend on the finite width of the cell only, but also on the number and type of particles and the volume of liquid available above the granular bed. The influence of these parameters is presented in the Appendix showing typical snapshots of the stationary state when changing the granular bed height $h_g$, the total height of the liquid $h_\ell$ and the batch of particles. It can be seen that both $h_\ell$ and $h_g$ have an effect on the intensity, and therefore, the \rep{density}{solid fraction} of the suspension (see figure~\ref{fig:snapshothghl} in the Appendix). However, no clear trend can be highlighted.  Finally, as expected, the small particles (PS80~M) are more easily put into suspension than larger particles (\rep{PS~230M}{PS~250M}) (see figure~\ref{fig:snapshotbatch} in the Appendix).

\subsection{Temporal evolution and quantitative measurement of the granular bed size}

\begin{figure}
\includegraphics[width=0.8\linewidth]{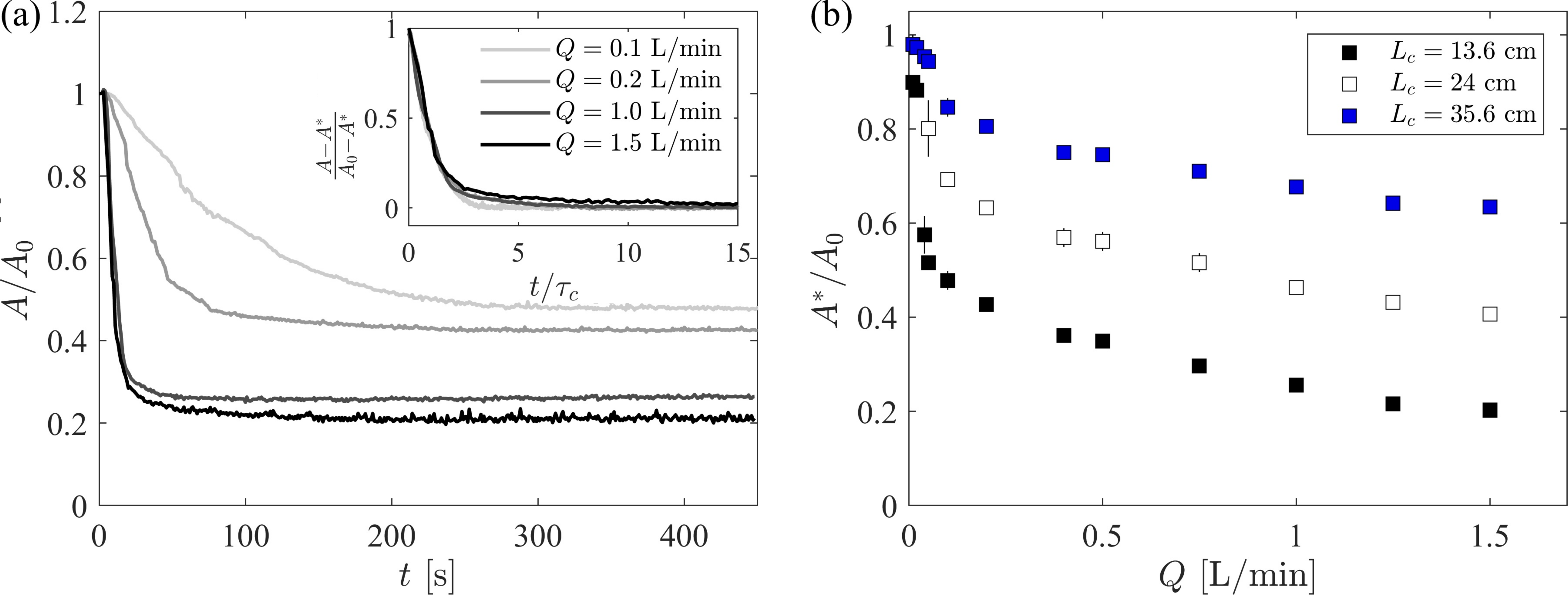}
\caption{  \label{fig:TempEvoAPhi}
(a) Temporal evolution of the \rep{dead zone}{bed} normalized area, $A/A_0$ for different air flow-rate $Q$ (decreasing for light gray to dark gray). Experimental parameters are those given in the caption of figure~\ref{fig:LcQdiagram}  [PVC 110P, $h_g = 9\pm 0.5$~cm, $h_\ell = 18\pm0.5$~cm\add{, $L_{\textrm{c}}=13.6$~cm, $e=2$~mm}].
\add{Inset: Normalized plot, $(A-A^*)/(A-A_0)$, as a function of $t/\tau_c$.} (b) \rep{Dead zone}{Bed} normalized area in the stationary state, $A^*/A_0$, as a function of the injected air flow-rate $Q$ for different cell width. \add{[$e=2$~mm for $L_c=13.6$~cm and $24$~cm and $e=3$~mm for $L_c=35.6$~cm]}}
\end{figure}

Figure~\ref{fig:TempEvoAPhi}(a) displays the temporal evolution of the normalized bed volume (or area), $A/A_0$, defined as its surface $A$ times the cell gap $e$, relative to the initial bed volume, $A_0e$. For different air injection flow rates $Q$ (increasing from $0.1$~L/min to $1.5$~L/min), the system always reaches a stationary regime in which the volume of the granular bed remains constant. \add{The final area of the bed is denoted $A^*$.} As expected, the characteristic time $\tau_c$ to reach the steady state for $A/A_0$ decreases when increasing $Q$. Such behavior was previously reported for the fluidization of heavy particles bed ($d\sim 3$~mm, $\rho_g = 2230$~kg/m$^3$) in two-phase systems~\cite{Philippe2013}. \add{The normalized plot $(A-A^*)/(A-A_0)$ as a function of $t/\tau_c$ is displayed as an inset of Figure~\ref{fig:TempEvoAPhi}(a) and shows that all the curves collapse on the same master curve, indicating that the dependence of $\tau_c$ with $Q$ is related to the dependence of $A^*$ with $Q$.} In the present paper, we focus only on the characteristics of the granular bed and the suspension in the stationary regime. In the following, the quantities in the stationary state are denoted with an asterisk.

To quantity the phenomenological observations, the final volume of the granular bed is displayed in figure~\ref{fig:TempEvoAPhi}(b) as a function of the flow rate for different values of the cell width $L_c$. For each cell width, $A^*/A_0$ decreases with $Q$. It drops abruptly at small flow rates $Q<250$~mL/min, while for $Q>250$~mL/min, the size of the \rep{dead zone}{granular bed} decreases gently.  In addition, the decrease of $A^*/A_0$ as a function of $Q$ is more abrupt for small cell widths. Thus, more particles remain in the granular bed when increasing the flow-rate for large $L_c$ compared to small $L_c$. This result can be easily explained, since for large $L_c$ most particles are far from the injected point and a larger flow rate is necessary to reach them. In addition, when the particles are resuspended into the fluid, they have a larger volume they can occupy for a large cell that for a small cell. The \rep{packing fraction}{solid fraction} of the induced suspension is therefore also smaller for large $L_c$.

\DEL{Mean packing fraction of the suspension}
\subsection{Mean solid fraction of the suspension}
\label{sec:phis}

\begin{figure}[t!]
\includegraphics[width=0.85\columnwidth]{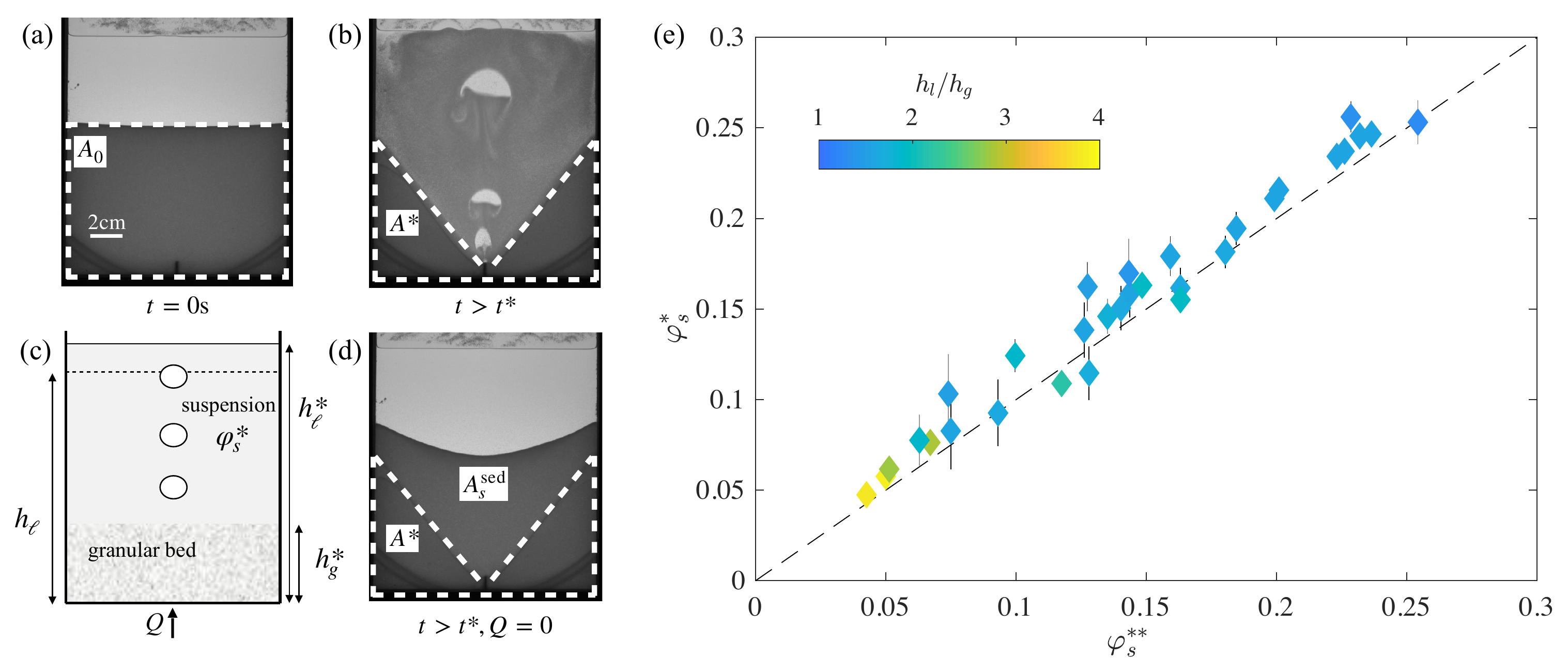}
\caption{ \label{fig:masscons}
Snapshots of the experiments just after turning on the flow rate (a) and in the stationary regime (b). [PS 130P; $Q=100$~mL/min; $L_c=13.6$~cm\add{; $e=2$~mm}; $h_g = 9.3\pm 0.5$~cm; $h_\ell = 14.4\pm0.5$~cm]
(c) Schematic view of the mass conservation model used to compute the \rep{packing fraction}{solid fraction},  of the suspension. (d) Snapshot of the experiments after turning off the flow rate.
(e) \rep{packing fraction}{Solid fraction} of the suspension $\varphi_s^*$, computed using the mass conservation model (eq.~\ref{eq:phis_mc}), as a function of  \rep{$\varphi_s^{\textrm{sed}}$}{$\varphi_s^{\textrm{**}}$}, computed using the number of particles that have sedimented after turning off the flow rate for different ratio $h_{\ell}/h_g$. The error bars correspond to the estimation of the compaction of the granular bed ($\phi_b^*=(1.05 \pm 0.05)\phi_b^0$). The dashed line corresponds to the first bisector.}
\end{figure}

As underlined in the introduction, an important quantity is the number of particles in the particle-laden liquid above the granular bed. In the previous section, we qualitatively comment on the \rep{packing fraction}{solid fraction} of the suspension. Here, we present a quantitative method to compute the mean \rep{packing fraction}{solid fraction} $\varphi_s^*$ of the suspension from the measurement of the final bed size $A^*/A_0$ using mass conservation. The total number of particles in the suspension, denoted $N_s^*$,  occupies a volume \rep{$V_s=A_s e$}{$V_s^*=A_s^* e$}, where \rep{$A_s$}{$A_s^*$} is the area occupied by the suspension. The mean \rep{packing fraction}{solid fraction} of the suspension can therefore be written as \rep{$\varphi_s^* = N_s^* V_g/V_s$}{$\varphi_s^* = N_s^* V_g/V_s^*$}, where $V_g \simeq (4/3)\pi(d/2)^3$ is the typical grain volume. The final bed size, $A^*/A_0$, and $\varphi_s^*$ are not independent variables, since they are directly linked by the particles mass conservation. Indeed, the number of grains $N_g$ is fixed in the experiment and can be computed using the bed area in the initial state (figure~\ref{fig:masscons}a). In the stationary state (figure~\ref{fig:masscons}b), it can be decomposed into two populations: $N_{\textrm{bed}}^{*}$ particles, which are still in the granular bed, and $N_s^*$ particles, which have been lifted in suspension, such that
\begin{equation}
N_g= N_{\textrm{bed}}^{*}+N_s^*\,.
\end{equation}
Using the definition of the \rep{packing fraction}{solid fraction}, one can get
\begin{equation}
\varphi_b^0 A_0 = \varphi_b^* A^* + \varphi_s^* \DEL{\textcolor{red}{V_s}}\add{A_s^*}\,,
\end{equation}
where $\varphi_b^*$ is the \rep{packing fraction}{solid fraction} of the granular bed in the stationary state. \rep{After a quick compaction by a few percent at the beginning of the experiment, the bed packing remains constant.}{At the beginning of the experiment, we observe a quick compaction by a few percent of the granular bed. This observation can be related to the compaction reported classically for dry or immersed granular beds under mechanical vibration, which are here triggered by the upward flow~\cite{Gauthier19}. After this quick compaction, the bed packing remains constant.} To get an estimate of the volume occupied by the suspension, a typical sketch of the stationary state regime is proposed in figure~\ref{fig:masscons}c). The liquid height in the stationary state $h_\ell^*$ is larger than in the initial condition due to the presence of the bubbles. Moreover, in most of the experimental configurations presented in this paper, the suspension has reached the side boundaries of the cell. \rep{, i.e.}{In the following, we consider only data for which} $L_s \simeq L_c$. The \rep{area}{volume} occupied by the suspension is therefore equal to $h_{\ell}^* L_c e - A^*e-V_{\textrm{Bubbles}}$, where the final \rep{area}{volume} of the granular bed, $A^*e$ and the volume of the bubbles, $V_{\textrm{Bubbles}}$ have been subtracted to the total \rep{area}{volume} $h_{\ell}^* L_c e$ occupied by the system. Since the liquid is incompressible, $h_{\ell}^* L_c e - V_{\textrm{Bubbles}}$ is simply equal to $h_{\ell} L_c e$. After some algebra, one gets the \rep{packing fraction}{solid fraction} in the suspension, in the stationary regime, as a function of $A^*/A_0$:
\begin{equation}
\varphi_s^*= \frac{\varphi_b^0 - (A^*/A_0) \varphi_b^*}{(h_\ell/h_g) - (A^*/A_0)}   \, .
\label{eq:phis_mc}
\end{equation}

This estimate can be compared to a more direct measurement of the number of particles in the suspension. At the end of the experiment, the air flow is turned off and all the particles in suspension sediment in loose packing granular bed with the same \rep{packing fraction}{solid fraction} of the initial state, $\varphi_b^0$. Since $\varphi_b^0$ and $\varphi_b^*$ are slightly different, two regions in this final granular state  can be distinguished (figure~\ref{fig:masscons}d): the granular bed of size $A^*$ and the bed formed by the particule previously in suspension of size $A_s^{\textrm{sed}}$. \DEL{The packing fraction in the suspension in the stationary state is equal to}\add{The solid fraction in the latter bed is equal to}
\begin{equation}
\varphi_b^0=\frac{N_s^* V_g}{A_s^{\textrm{sed}} e}\textcolor{red}{\DEL{\frac{N_s^* V_g}{V_s}}}
\label{eq:sedimentation_end}
\end{equation}
\add{Using Eq.~\ref{eq:sedimentation_end} and the expression of the volume occupied by the suspension as a function of $A^*$, o}ne can get a second estimate of the \rep{packing fraction}{solid fraction} of the suspension, denoted $\varphi_s^{\textrm{**}}$, as
\begin{equation}
\DEL{\varphi_s^{\textrm{sed}}}\add{\varphi_s^{\textrm{**}}}=\add{\frac{N_s^* V_g}{V_s^*}\;=}\;\varphi_b^0\frac{A_s^{\textrm{sed}}}{h_\ell L_c - A^*}\,.
\end{equation}

Figure~\ref{fig:masscons}(d) displays the \rep{packing fraction}{solid fraction} measured in the stationary state, $\varphi_s^{*}$ as a function of the one computed at the end of the experiment, \rep{$\varphi_s^{\textrm{sed}}$}{$\varphi_s^{\textrm{**}}$} for different values of the height ratio, $h_\ell/h_{\textrm{g}}$.  As expected, all the data points collapse on the first bisector within experimental error bars, showing that the two estimations are in very good agreement, which validates the computation of the suspension \rep{packing fraction}{solid fraction} using the bed final size. For all experiments, we can thus determine the average \rep{packing fraction}{solid fraction} in the suspension, $\varphi_s$, at all time. In the following sections, we focus on the stationary state, where this \rep{packing fraction}{solid fraction} is denoted $\varphi_s^*$.

\section{Phenomenological model}
\label{sec:model}

In this section, we propose a phenomenological model to explain the dependence of the spatial average \rep{packing fraction}{solid fraction} in the suspension, $\varphi_s$, on the experimental parameters. In the stationary state the suspension results from the balance between grains advected by the bubbles rising in the above liquid layer, and particles settling on the sides. 
The following subsections propose an expression of the number of particles sedimenting on the sides of the cell (sec.~\ref{sec:sedimentation}) or entrained by the bubbles at the centre (sec.~\ref{sec:entrainment}), with a final expression of the suspension average \rep{packing fraction}{solid fraction} in the stationary state, $\varphi_s^*$, resulting from the competition between both mechanisms (sec.~\ref{sec:balance}).

\subsection{Particle settling}
\label{sec:sedimentation}

\begin{figure}[t]
\includegraphics[width=0.6\linewidth]{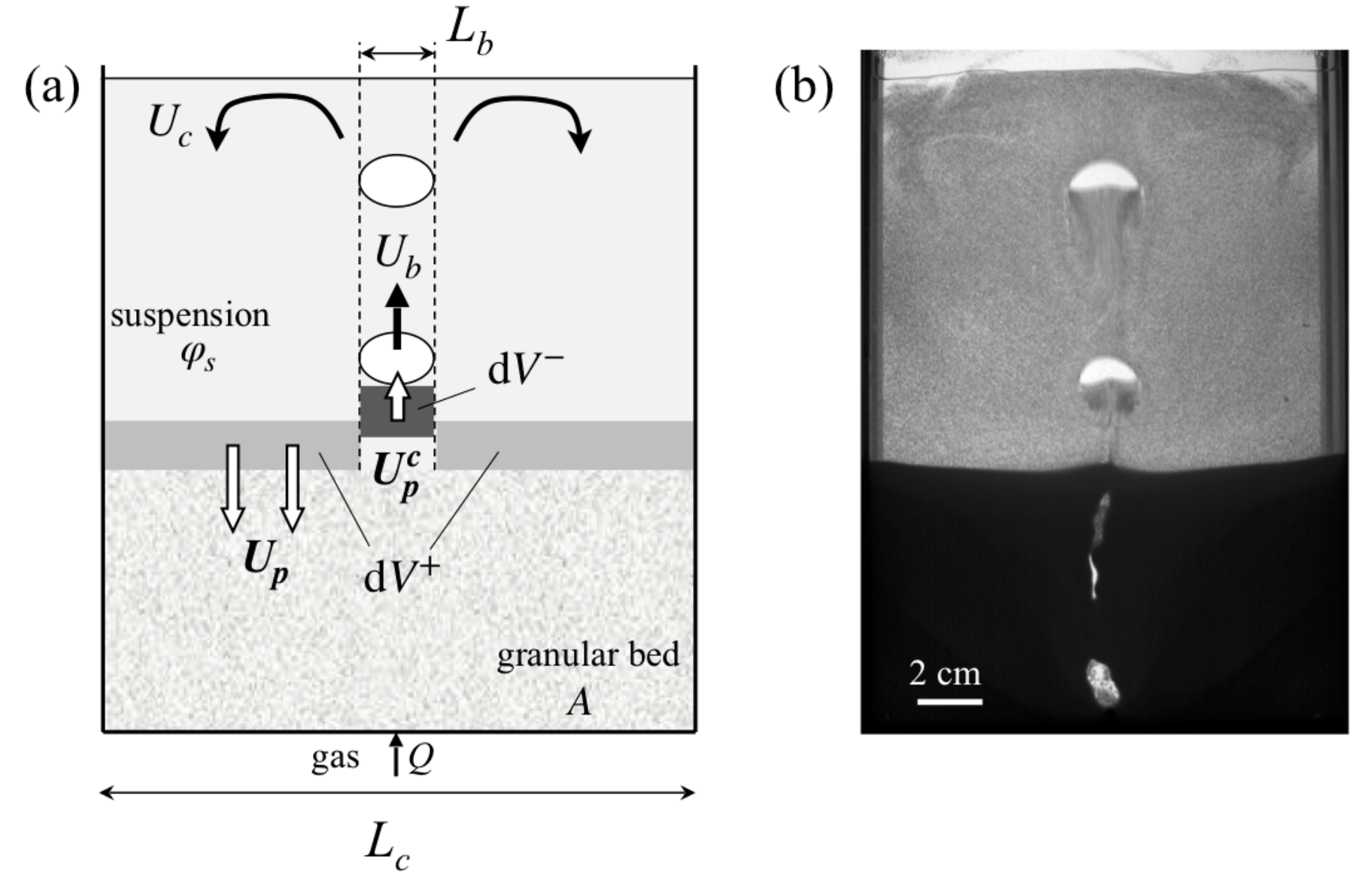}
\caption{
(a) Sketch of the fluid-particle suspension by gas injection (phenomenological model, see section~\ref{sec:model}).  Gas is injected at constant flow rate $Q$ at the bottom center of the cell and cross{es} the granular bed. Bubbles rising in the above liquid layer (velocity \rep{$U_B$}{$U_b$}) entrain particles upwards in the central column (width \rep{$L_B$}{$L_b$}) and form a suspension of width $L_S$ \add{which is of the order of $L_c$ in all the data discussed in section IV.} Particles on both sides settle due to both sedimentation and fluid recirculation (see text). 
(b) Picture of bubbles rising out of the granular bed entraining particles in their wake [PVC 110P, $h_g=9$~cm, $h_\ell=18$~cm, $Q=0.05$~L/min\add{; $L_{\textrm{c}}=13.6$~cm, $e=2$~mm}].
}
\label{fig:model} 
\end{figure}

Let us note $\mathrm{d}N^+$ the number of particles in a volume $\mathrm{d}V^+$ of the suspension settling on the granular bed during the time interval $\mathrm{d}t$ (Figure~\ref{fig:model}a, dark gray zones on the suspension sides),
\begin{equation}
\textrm{d}N^+ = \varphi_s \, \frac{ \textrm{d}V^+}{V_g} 
                        = \varphi_s \, \frac{ (L_s-L_b) e U_p \textrm{d}t}{V_g}\,,
\end{equation}
where $V_g$ is the volume of a single grain, $\varphi_s$ the \rep{packing fraction}{solid fraction} of the suspension, and $U_p$ the particle settling velocity. As a reminder, $L_s$ represents the typical width occupied by the suspension, which reaches the lateral boundaries of the cell in \rep{most of the experiments  presented in this paper}{all the data that will be discussed in this model}, so that $L_s\approx L_c$. \DEL{Following Maxey \& Riley~\cite{Maxey1983}} \add{The particle sedimentation velocity $U_p$ can be expressed using the relative velocity between a particle and the surrounding suspension which is often written as~\cite{Lhuillier2009,Nott2011,Guazzelli2018}
\begin{equation}
U_p-U=U_s(1-\varphi_s)^5\,.
\end{equation}}
\DEL{The particle sedimentation velocity $U_p$ can be expressed as the sum of the particle settling velocity in the quiescent fluid and of the average velocity of the \rep{suspension}{surrounding liquid $U_f$} on the cell sides}
\begin{equation}
\DEL{U_p = U_s^* + U_f\,.}
\end{equation}
\add{Since the  Archimedes number Ar, which corresponds to the particle Reynolds number based on the Stokes velocity $U_s=\Delta \rho g d^2/(18 \mu)$, is smaller than 1 (see Table~\ref{tab:grains}), the Stokes velocity $U_s$ is the pertinent settling velocity, and the term $(1-\varphi_s)^5$ corresponds to the correction due to collective effects~\cite{Richardson1954,Guazzelli2012}. The velocity $U$ corresponds to the velocity of the suspension, which is the volume average velocity, and takes into account the velocity of the particles $U_p$ and the velocity of the fluid $U_c$ due to the recirculation generated by the rising bubbles:
\begin{equation}
U = \varphi_s U_p + (1-\varphi_s) U_c\,.
\end{equation}
Using mass conservation, the average fluid velocity given by the recirculating flow is $U_c = U_b L_b / (L_s-L_b)$.
}
%
\DEL{Combining the above equations leads to }\add{The particule velocity on the cell sides can therefore be written $U_p = U_c + U_s (1-\phi_s)^4$ and leads to an expression for the number of particles settling on the granular bed during $\mathrm{d}t$:}
\begin{equation}
\textrm{d}N^+ = \varphi_s \, \frac{ (L_s-L_b) e}{V_g} 
                            \left[\DEL{\textcolor{red}{{U_S}}}\add{U_s}(1-\varphi_s)^4 + \frac{U_b L_b}{L_s-L_b}   \right]  \textrm{d}t   \, .
\end{equation}

\subsection{Entrainment}
\label{sec:entrainment}

Let us denote $\mathrm{d}N^-$ the number of particles entrained during the \DEL{the} same time interval $\mathrm{d}t$ (Figure~\ref{fig:model}a, black zone in the central column). The particles entrained can come either from the granular bed or from the recirculating suspension. We thus write:
\begin{equation}
\textrm{d}N^- = (\chi_s \varphi_s + \chi_b \varphi_b) \, \frac{ L_b e U_p^c \textrm{d}t }{V_g} 
\end{equation}
where $\chi_s$ and $\chi_b$ are coefficients representing the fraction of particles entrained from the suspension, $\chi_s$, or the granular bed, $\chi_b$, and $U_p^c$ is the particle velocity in the bubble's wake\DEL{,}\add{. The latter can be written, as previously for the particles on the cell sides, as the velocity composition between the fluid, equal to the bubble rising velocity $U_b$ in the central column, and the term due to particle sedimentation:
\begin{equation}
U_p^c-U^c=U_s(1-\varphi_s)^5\,,
\end{equation}
in which $U^c$ is the velocity of the suspension in the central region and is given by an average between the velocity of the particle $U_p^c$ and the fluid velocity, equal to the bubble velocity $U_b$:
\begin{equation}
U^c = \varphi_s U_p^c + (1-\varphi_s) U_b\,.
\end{equation}
The combination of the two previous equations leads to}
\begin{equation}
U_p^c = U_b - \DEL{U_S}\add{U_s} (1-\varphi_s)^4  .
\end{equation}
The number of particles entrained during a time inverval $\textrm{d}t$ is therefore 
\begin{equation}
\textrm{d}N^- = (\chi_s \varphi_s + \chi_b \varphi_b) \, \frac{ L_b e}{V_g}
                           \left[U_b - \DEL{U_S}\add{U_s}(1-\varphi_s)^4  \right]  \textrm{d}t   \, .
\end{equation}

\subsection{Stationary state}
\label{sec:balance}

In the stationary state, $\textrm{d}N^+ = \textrm{d}N^-$. We introduce the dimensionless variables $\ell=L_b/L_s$ and $u = U_b/\DEL{U_S}\add{U_s}$. \add{Note that the parameter $u$ represents the inverse of a Rouse number defined as the velocity ratio between the entraining, rising fluid and the sedimentation of a single particle.} After some algebra, one gets the following relation for $\varphi_s^*$, the \rep{packing fraction}{solid fraction} in the suspension in the stationary regime:
\begin{eqnarray}
\varphi_s^* = (\chi_s \varphi_s^* + \chi_b \varphi_b^*) \,   
                   \frac{u- (1-\varphi_s^*)^4}{u + \frac{1-\ell}{\ell} (1-\varphi_s^*)^4}\,.
                   \label{eq:model_steady}
\end{eqnarray}

\section{Particle suspensions}
\label{sec:Particle_susp}

In this section, we compare the average \rep{packing fraction}{solid fraction} of the suspension inferred from experimental measurements of $A^*/A_0$, in the stationary regime (equation~\ref{eq:phis_mc}), with the model prediction of its variation upon $u,\ell$ and $\varphi_b^*$ (equation~\ref{eq:model_steady}). In section~\ref{sec:LbUb}, we propose an estimation of the bubble typical size and velocity.
$\chi_s$ and $\chi_b$ are {\it a priori} unknown and will be discussed in section~\ref{sec:mechanism}.

\subsection{Typical bubble size and velocity}
\label{sec:LbUb}

\begin{figure}[t!]
\includegraphics[width=1\columnwidth]{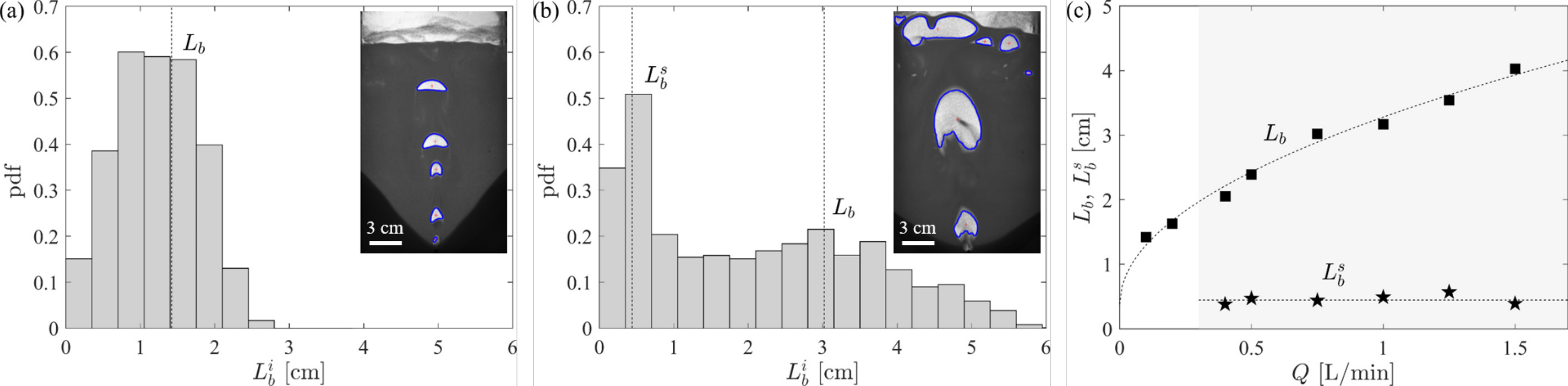}
\caption{ \label{fig:bubblediam}
(a,b) Probability density function of the bubble equivalent diameter ($L_b^i$ indicates the equivalent diameter for bubble $i$) in the stationary regime. The inset displays a snapshot of the corresponding image sequence with bubble contour (blue line) and center of mass (red cross). (a) At low flow rate [$Q=0.1$~L/min], the bubble population displays a single characteristic size $L_b$. (b) At high flow rate [$Q=0.75$~L/min], small bubbles resulting from bubble fragmentation appear, with a typical size $L_b^s$.
(c) Equivalent bubble diameter as a function of the flow rate $Q$. For $Q>0.3$~L/min, we observe two bubble populations (corresponding to the two peaks in (b)). Small bubbles are roughly constant in size (horizontal dashed line) while large bubbles follow $L_b = 0.38+2.9\sqrt{Q}$ (increasing dashed line).
[PVC~110P; $L_c=13.6$~cm\add{; $e=2$~mm}; $h_g = 9.3\pm 0.5$~cm; $h_\ell = 18\pm0.5$~cm].
}
\end{figure}

In confined geometry, previous studies have shown that the thickness $e_f$ of the lubrication film between the bubble and the wall is given by $e_f/e \sim \mathrm{Ca}^{2/3} / (1+\mathrm{Ca}^{2/3})$, where $\mathrm{Ca}=\mu U_b/\sigma$ is the capillary number \cite{Aussillous2000,Roudet2008} and $\sigma \simeq 22$~mN/m the air/ethanol surface tension at room temperature. In our experiments\add{, a rough estimate of the bubble velocity is} $U_b \sim 10$~cm/s, leading to $e_f/e \ll 1$. We can therefore estimate the bubble volume as their apparent surface $S$, computed from the images, multiplied by the cell gap $e$. The typical bubble size can then be estimated as its equivalent diameter, $L_b=\sqrt{4S/\pi}$.

Figures~\ref{fig:bubblediam}(a) and~\ref{fig:bubblediam}(b) display histograms of bubble population in the presence of PVC particles, at two different flow rates, in the stationary regime. The picture in inset of each figure shows an example of bubble contour detection (in blue). For $Q>0.3$~L/min, a population of small bubbles appear jointly with the larger bubbles. The maxima of each distribution is picked and reported in figure~\ref{fig:bubblediam}(c). The small bubble equivalent diameter, $L_b^s$, remains roughly constant and of the order of 4.5~mm as a function of the gas flow-rate $Q$. They are generated by bubble fragmentation and almost always on the sides of the central vertical line above the injection nozzle. Consequently, they do not play any significant role in particle entrainment in the central zone, and will be further ignored.
The larger bubble size \rep{increase}{increases} as the square root of the flow-rate, here $L_b = 0.38+2.9\sqrt{Q}$ for the PVC~110P. This dependence on $\sqrt{Q}$ does not seem to vary significantly when changing the particles\add{, and is not directly governed by gravity either, as pointed out by previous experiments with PVC particles in tilted cells~\cite{PicardRNL2018}.}

\add{The bubble velocity $U_b$ cannot be measured directly in our experiments. Indeed, we capture the stationary state of the suspension over long times, and the acquisition frame rate (1~Hz) is too low to determine the bubble rising speed. From rough observations, we can estimate the bubble velocity between a few and a few tens of centimeters per second.}
\add{It is of the same order of magnitude as the typical rise time of bubbles in a Hele-Shaw cell filled with water only. Previous works experimentally investigated the velocity of bubbles rising in Hele-Shaw cells and found that, for Reynolds number smaller than $10^3$, as in our experiments, the bubble velocity can be written as $U_b\sim 0.5 \sqrt{g L_b}$~\cite{Roudet2008,Roig2012}. This expression can be interpreted as a simple balance between the buoyancy force and the drag force excerted on the free edges of the bubble, the contribution of the viscous shear stress of the liquid films being negligible~\cite{Roudet2008,Roig2012,Filella2015}.}

\subsection{Entrainment mechanism}
\label{sec:mechanism}

\begin{figure}
\includegraphics[width=0.75\linewidth]{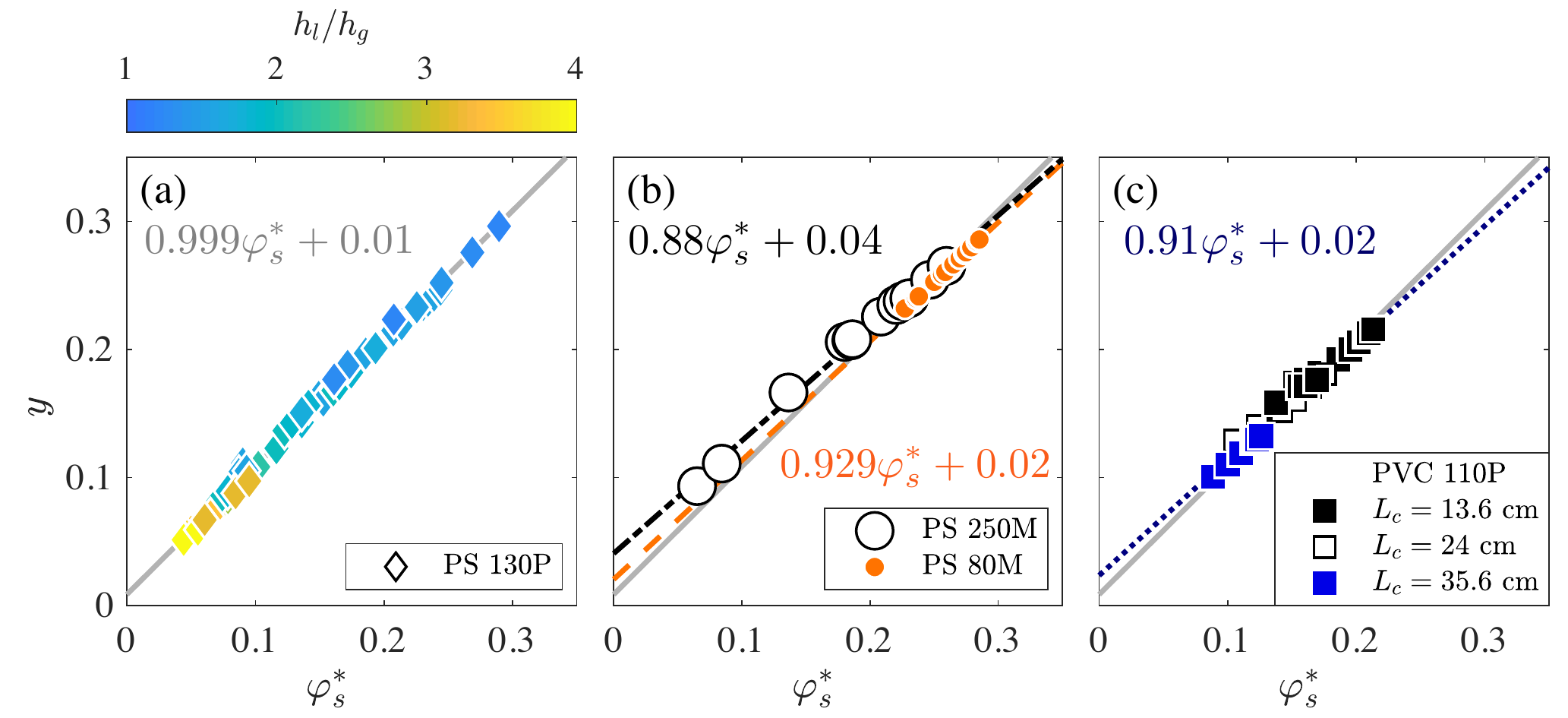}
\caption{Parameter $y$ as a function of $\varphi_s^*$ (see text). Following the prediction of the phenomenological model, $y=\chi_s \varphi_s^* + \chi_b \varphi_b^*$ where $\chi_s$ and $\chi_b$ depend on the nature of the particles only. 
(a) PS~130P for different $h_\ell/h_g$ (cf. colorscale) [$L_c=13.6$~cm\add{, $e=2$~mm}]. The solid gray line indicates the linear fit (reported in (b) and (c) for comparison)\add{, whose equation is indicated in gray on the figure.}
(b) PS~250M and PS~80M [$h_g=9$~cm, $h_\ell=18$~cm, $L_c=13.6$~cm\add{, $e=2$~mm}]. Dashed black (resp. orange) line: linear fit for the PS~250M (resp. PS~80M) particles. \add{The fit equations are indicated in black (resp. orange) in the figure.}
(c) PVC~110P [$h_g=9$~cm, $h_\ell=16$~cm]. All data collapse independently of the cell width $L_c$ (blue dotted line: linear fit, \add{whose equation is indicated in blue}). \add{[$e=2$~mm for $L_c=13.6$~cm and $24$~cm and $e=3$~mm for $L_c=35.6$~cm]}}
\label{fig:chismodel} 
\end{figure}

In figure~\ref{fig:chismodel}, we plot $y=\varphi_s^* / \left[  \frac{u- (1-\varphi_s^*)^4}{u + \frac{1-\ell}{\ell} (1-\varphi_s^*)^4}  \right]$ as a function of $\varphi_s^*$. \add{Our experiments have a range of the two parameters $\ell$ and $u$ of $0.019 < \ell < 0.31$ and $24 < u < 427$ respectively.}  As predicted by the phenomenological model, for a given type of particles, all data collapse on a linear trend, $\chi_s \varphi_s^* + \chi_b \varphi_b^*$. This result is independent of the height ratio $h_\ell/h_g$ (Figure~\ref{fig:chismodel}a) or of the cell width (Figure~\ref{fig:chismodel}c), as long as $L_s \simeq L_c$ (see section~\ref{sec:model}). \add{The different coefficients of the linear trend depend on the different batches of grains as indicated by the fit equations provided for all data sets in Figure~\ref{fig:chismodel}.} The model therefore predicts well, at first order, the behavior of the suspension. Interestingly, for all particles, we find \rep{$\chi_s \ll \chi_b$}{$\chi_s \gg \chi_b$}, with $\chi_s$ of the order of 90--99~\% and $\chi_b$ of the order of 2--7\%. This means that, in the stationary regime, the majority of the particles forming the suspension come from the global recirculation, and not from particles that have settled in the bed and are later extracted and entrained by the bubbles. PS~250M (black dotted line in figure~~\ref{fig:chismodel}b) and PS~130P (gray thick line in figures~\ref{fig:chismodel}a,b) clearly follow a different linear trend. Several features can explain this difference: the particle different size (see Table~\ref{tab:grains}), shape (angular vs. spherical, see figure~\ref{fig:expsetup}b,c) or polydispersity (monodisperse PS~250M vs. polydisperse PS~130P, see Table~\ref{tab:grains}). Figure~\ref{fig:chismodel}b compares two different sizes of the same monodisperse, polystyrene particles, PS~250M and PS~80M. Although a difference in the linear trend seems to appear, with PS~80M being closer to PS~130P particles trend (gray solid line, figure~\ref{fig:chismodel}b), it is difficult at this point to conclude unequivocally. Indeed, the PS~80M particles are easily entrained and sediment very slowly (see Stokes velocity, Table~\ref{tab:grains}) and it was impossible, in our experimental conditions, to form suspensions with $\varphi_s^* < 20$\%. Similarly, although the linear trends characterizing the PS~130P and PVC~110P particles appear slightly different (figure~\ref{fig:chismodel}c), they cannot be distinguished unambiguously due to the difficulty to span a large range of $\varphi_s^*$ for PVS~110P.
Therefore, within the experimental error bars, we cannot conclude with the present work on the dependence of the parameters $(\chi_s,\chi_b)$ on the particle sedimentation velocity or polydispersity.

\section{Conclusion}\label{sec:conclusion}
{In this paper, we have explored experimentally the resuspension of particles of an initially loosely packed immersed granular bed. We have observed that continuous air injection at the bottom of the granular bed leads to a final steady state for different sets of controlling parameters. This final steady state consists of a crater-shaped final granular bed  and a more or less homogeneous suspension formed by the particles entrained in the above liquid layer by the gas bubbles emerging from the granular bed. The final global characteristics of the suspension are quantified  by the spatial-averaged \rep{packing fraction}{solid fraction}, which is computed using the final granular bed volume based on mass conservation. Finally, we have finally proposed a phenomenological model for the steady state, reflecting the balance between the entrainment of particles by the air bubbles and their settling on the sides of the experimental cell. This model captures with good agreement the main features of this resuspension mechanism using empirical laws for the gas bubbles size and velocity. Moreover, we have shown that the suspension is almost ``self-sustained", meaning that almost all the particles entrained by the air bubbles come from a global recirculation mechanism and therefore do not go back into the granular bed. Only a few percent of particles are extracted from the granular bed.}

{Even if the phenomenological model gives interesting insights into the behavior of suspension generated by gas release, it cannot be predictive. Indeed, it strongly depends on the behavior of the gas bubbles, which depends in particular on the suspension itself. As underlined in~\cite{Dollet2018}, bubbles are still a challenge for scientists to understand and/or control their behavior in many complex situations. The effect of the suspension on the dynamics of bubbles is beyond the scope of this paper.} {It would also be interesting to study not only the global behavior of the suspension but also the local evolution of the \rep{packing fraction}{solid fraction} since the inhomogeneities may be very large. These perspectives shall be the topic of future studies.}


\section*{Acknowledgements}
{We thank D. Bén\^atre for insightful discussions. This work was supported by the LABEX iMUST (ANR-10-LABX-0064) of Université de Lyon, within the program "Investissements d'Avenir" (ANR-11-IDEX-0007) operated by the French National Research Agency (ANR). The authors thank two anonymous referees for comments which greatly improved this work.}



\newpage
\appendix
\section{Stationary state\label{appendix_snapshots_parameters}}
\begin{figure}[h]
\includegraphics[width=0.6\linewidth]{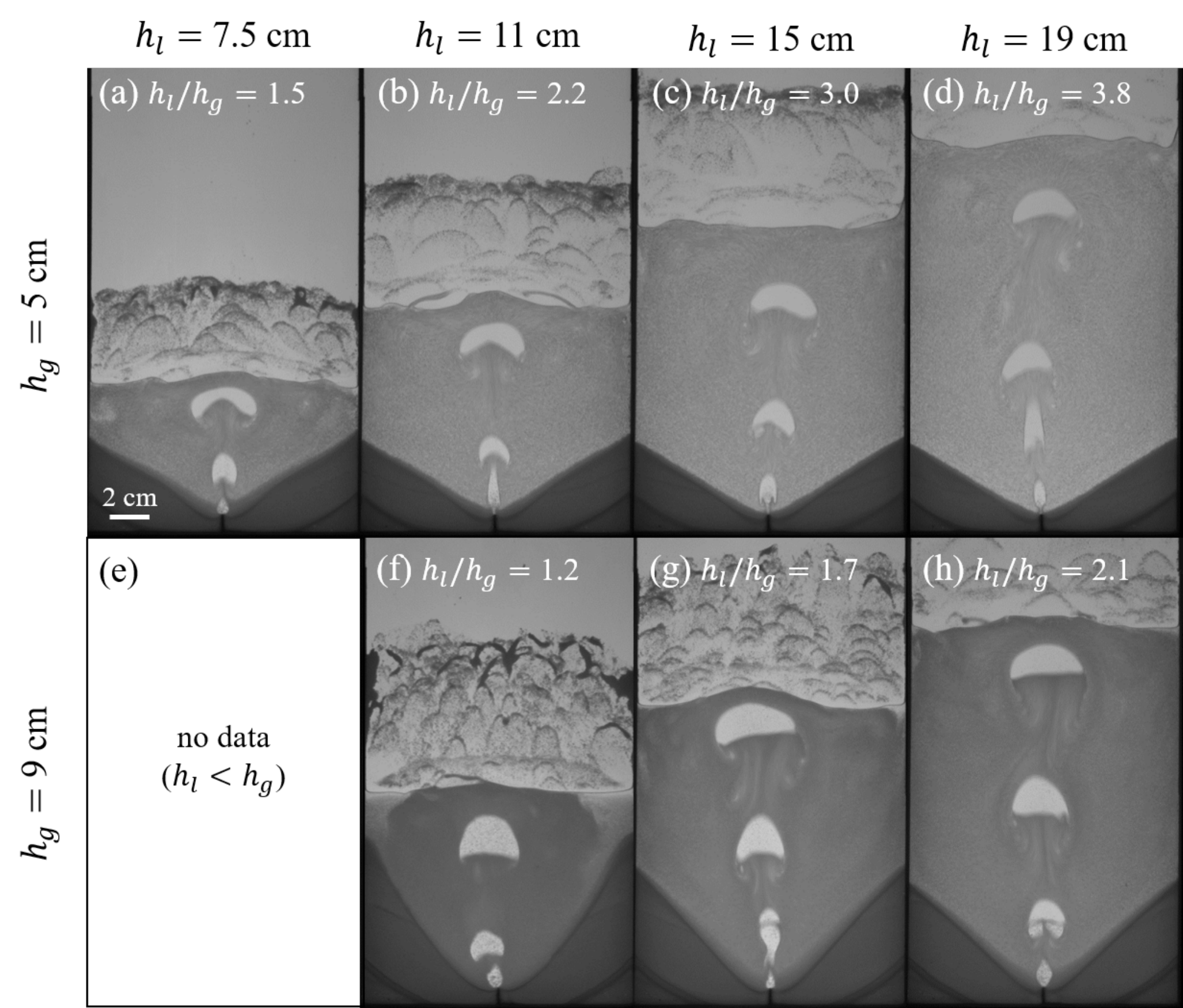}
\caption{ \label{fig:snapshothghl} Snapshots of the experiments in the stationary regime for different granular bed heights $h_g$ and liquid heights $h_\ell$ [PS 130P; $L_c = 13.6$~cm\add{; $e=2$~mm}; $Q = 200$~mL/min].}
\end{figure}

\begin{figure}[h]
\includegraphics[width=0.8\linewidth]{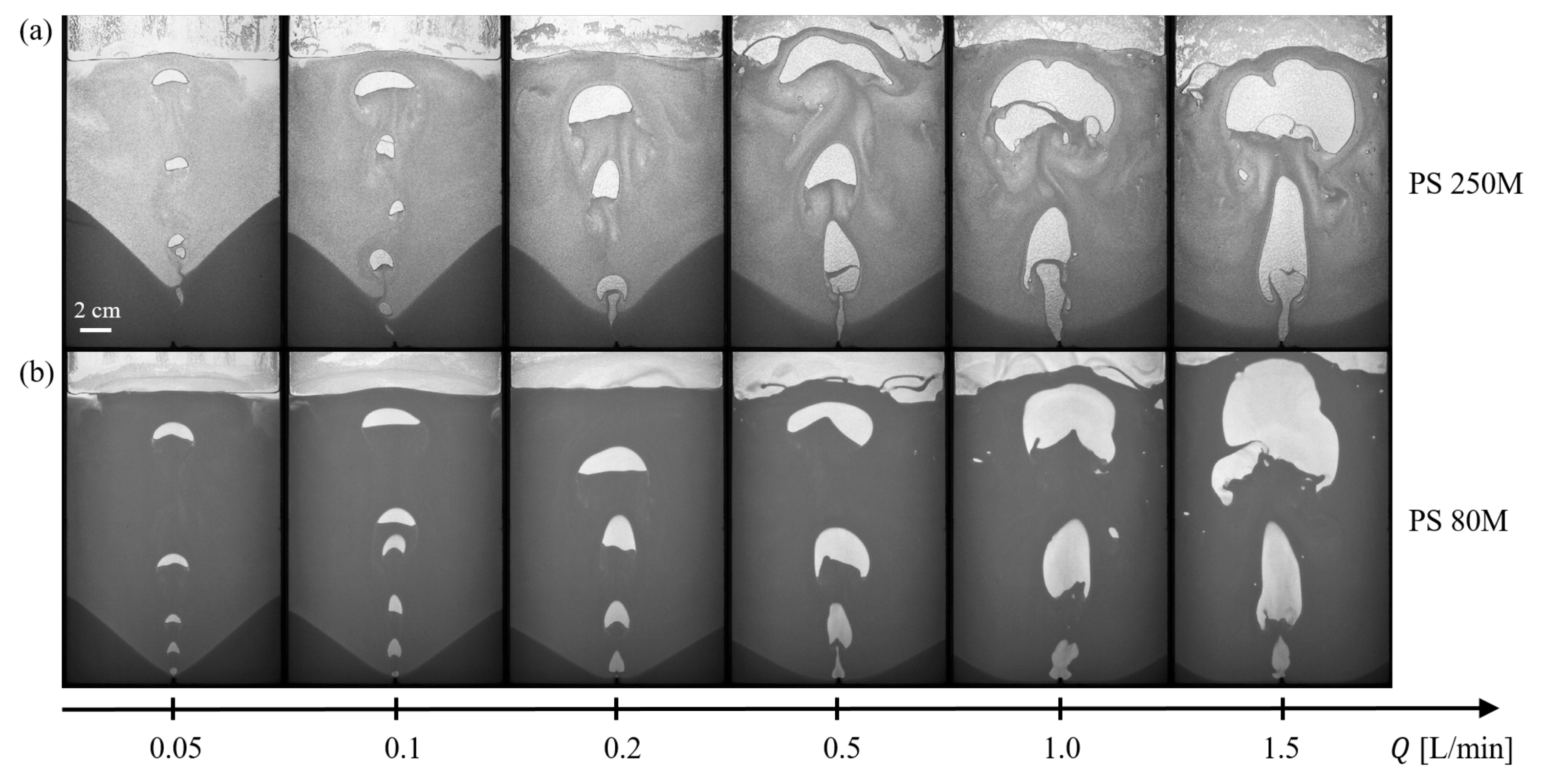}
\caption{ \label{fig:snapshotbatch}
Snapshots of the experiments in the stationary regime for different values of the flow rate $Q$ and the two different batches of monodisperse particles [$L_c=13.6$~cm\add{; $e=2$~mm}; $h_g = 9\pm 0.5$~cm; $h_\ell = 18\pm0.5$~cm].}
\end{figure}

In this Appendix, we show typical snapshots of the final steady states displayed in a similar way as in Fig~3 for additional sets of external parameters.
\begin{itemize}
\item Fig.~\ref{fig:snapshothghl} shows the influence of the granular bed height $h_g$ and the liquid height $h_\ell$. For each snapshot, the ratio $h_\ell/h_g$ is given, since it is the important reduced parameter for the computation of the packing fraction of the suspension (see Eq.~(3)). These snapshots show an effect of both parameters on the final granular bed area and the final packing fraction of the suspension, but they do not reveal any clear trend.
\item Fig.~\ref{fig:snapshotbatch} shows the influence of the flow rate $Q$ and the two different batches of monodisperse particles. As expected, the small particles (PS80~M) are more easily put into suspension than larger particles (PS~250M).
\end{itemize}

\end{document}